\documentclass[reprint, amssymb, amsmath, aip,cha, prb]{revtex4-2}

\usepackage{graphicx}   
\usepackage{amsmath, amssymb} 
\usepackage{chemformula} 
\usepackage{xcolor} 
\usepackage{siunitx} 
\usepackage{mhchem}  

\begin{document}

\title{Magnetic Circular Dichroism at the Oxygen K edge in Microcrystals of Spinels Grown on Ru(0001)}

\author{A. Mandziak}
\affiliation{Solaris Synchrotron, Czerwone, 30-392 Cracow, Poland}
\author{V. Sosa}
\affiliation{Instituto de Ciencia de Materiales de Madrid, CSIC, Madrid E-28049, Spain}
\author{P. Nita}
\affiliation{Solaris Synchrotron, Czerwone, 30-392 Cracow, Poland}
\affiliation{Faculty of Physics, Astronomy and Applied Computer Science, Jagiellonian University, 30-348 Cracow, Poland}
\author{L. Mart\'{\i}n-Garc\'{\i}a}
\affiliation{Instituto de Qu\'{\i}mica F\'{\i}sica Blas Cabrera, CSIC, Madrid E-28006, Spain}
\author{J. E. Prieto}
\affiliation{Instituto de Qu\'{\i}mica F\'{\i}sica Blas Cabrera, CSIC, Madrid E-28006, Spain}
\author{M. Foerster}
\affiliation{Alba Synchrotron Light Facility, CELLS, Barcelona E-08290, Spain}
\author{M. A. Niño}
\affiliation{Alba Synchrotron Light Facility, CELLS, Barcelona E-08290, Spain}
\author{L. Aballe}
\affiliation{Alba Synchrotron Light Facility, CELLS, Barcelona E-08290, Spain}
\author{C. Granados-Miralles}
\author{A. Quesada}
\affiliation{Instituto de Cer\'{a}mica y Vidrio, CSIC, Madrid E-28049, Spain}
\author{C. Tejera-Centeno}
\affiliation{Instituto de Ciencia de Materiales de Madrid, CSIC, Madrid E-28049, Spain}
\author{S. Gallego}
\affiliation{Instituto de Ciencia de Materiales de Madrid, CSIC, Madrid E-28049, Spain}
\author{J. de la Figuera}
\affiliation{Instituto de Qu\'{\i}mica F\'{\i}sica Blas Cabrera, CSIC, Madrid E-28006, Spain}

\date{\today}

\begin{abstract}

We have measured the circular magnetic dichroism in the x-ray absorption at the K-edge of oxygen in microcrystals of different spinel oxides. The microcrystals are islands of micrometric size and nanometric thickness, grown on Ru(0001) substrates using high-temperature oxygen-assisted molecular beam epitaxy. The domains observed in the oxygen K-edge dichroism have the same distribution and orientation as those
observed in x-ray magnetic circular dichroism at the L$_{3}$ edge of the octahedral cations. Integrating the area from a single domain, x-ray magnetic circular dichroic spectra of oxygen were measured and, by
the application of the K-edge sum rule, non vanishing orbital magnetic moments aligned with the octahedral cations were found. Density functional theory calculations, which did not show any orbital moment at the oxygen
anions, indicate that the energy ranges where oxygen dichroism is observed correspond to those with significant hybridization with the cations d bands. They also show a correlation between the magnitude of the measured
value of the oxygen orbital moment and the theoretical one for the cations, and demonstrate that this trend is preserved in the presence of Fe excess in the samples. Our experimental XMCD suggest, following the DFT calculations, that the origin of the oxygen magnetic moment lies in the hybridization of the oxygen unoccupied p-derived bands with the cation bands, mostly with the d-derived ones.
\end{abstract}

\maketitle

\section{Introduction}

Transition-metal oxides\cite{brabers_progress_1995} possess intriguing physical properties that make them both fascinating for fundamental research and interesting for various technological applications. These properties arise from the interplay between charge, orbital character and spin of the valence electrons, and the lattice dynamics. In particular, ferrimagnetic cobalt and nickel spinel ferrites have garnered significant interest for magnetic\cite{brabers_progress_1995} and spintronic applications\cite{LudersAM2006,bibes_oxide_2007,mesoraca_growth_2018}, due to their relatively high Curie temperatures and large stability and unique combination of distinct magnetic features\cite{Tejera2021}. A challenging aspect is the controlled growth of high quality microcrystals and films, due to the voids in the crystal lattice inherent to the spinel structure and the existence of multiple cation coordination sites. In their usual inverse form, the octahedral (B) sites in these materials are occupied by both divalent (Co$^{2+}$ and Ni$^{2+}$) and trivalent (Fe$^{3+}$) cations, while the tetrahedral sites (A) are populated by trivalent (Fe$^{3+}$) cations. However, while cobalt ferrite is mostly an inverse spinel with Co$^{2+}$ in the octahedral sites, depending on the growth method and the thermal history of the sample there can be a substantial amount of cobalt cations in tetrahedral positions\cite{de_la_figuera_mossbauer_2015}, with significant impact in the reduction of the magnetic anisotropy. Instead, nickel ferrite is typically mostly inverse. Such tendency is consistent with the existence of the Co$_3$O$_4$ spinel phase, while there is no known bulk Ni$_3$O$_4$ phase.

Experimentally, a very powerful technique to monitor the element resolved magnetic properties in spinels is through the study of the magnetic circular dichroism in the x-ray absorption of core
levels\cite{schutz_synchrotron_2007} (XMCD-XAS). In particular, XMCD spectra, i.e. the difference of the spectra acquired with right-hand and left hand circularly polarized x-rays, are very sensitive to the
magnetic moment of the corresponding atom. For the 3d-transition metal cations, the L$_{3,2}$ absorption edges in the soft x-ray region (700-900 eV) are usually employed. The integrals of the XMCD and the XAS spectra enable through the use of the sum rules to obtain the spin and orbital components of the element under study. In this way it has been detected that often the orbital magnetic moment of the cations in spinels is substantially reduced due to crystal field effects\cite{goering_large_2011,LauraPRB2015}. However, some 3d transition-metal cations exhibit sizable orbital magnetic moments, such as Co in cobalt ferrite\cite{kita_x-ray_2001,mund_investigation_2011,moyer_magnetic_2011,LauraAdvMat2015}. This phenomenon primarily arises from the asymmetric occupancy of the t$_{2g}$ orbitals of Co$^{2+}$ cations at the octahedral sites.

In this work we focus on the observation of moments not on the cations of spinels, but on the oxygen anions. Oxygen plays a primordial role in these materials: the octahedral and tetrahedral cations are far away enough
for the absence of any significant overlap between their orbitals. Instead, the p-derived bands of oxygen act through superexchange to promote the antiferromagnetic coupling between the d-derived bands of the cations
of both sublattices. Typically bands with p-orbital character have not been traditionally considered in magnetic systems, which rely instead on either delocalized d-electrons (3d transition metal elements such as Fe, Co and Ni), or on localized f-electrons (rare earth elements such as Gd, Tb or Sm) of the cations. Nevertheless, recent work, both experimental and theoretical, has highlighted the existence of the so-called p-magnetism in some oxides, prone to emerge when there are p-derived bands near the Fermi level\cite{pmag2005}. The oxygen K-edge, i.e. the excitation of electrons from the 1s core level to the unoccupied p-derived bands, is in the soft x-ray region, near 520 eV\cite{frati_oxygen_2020}. K-edge circular dichroism can also be used to estimate a magnetic moment, but due to the symmetry of the core level, only the magnetic orbital moment can be sampled\cite{guo_interpretation_1998,schutz_synchrotron_2007}. K-edges of cations in the hard x-ray region have been used to study cation's orbital moment and the application of the sum
rule in such cases is well established\cite{mathon_xmcd_2004,efimov_co_2016}. However, there is an ongoing discussion about the real meaning of the orbital magnetic moments obtained from the sum rule in the case of the
K-edge of light elements: unexpected large values are routinely obtained in different materials. An example that has been discussed at length is the case of the ferrimagnetic half-metal CrO$_2$\cite{goering_direct_2002,huang_orbital_2002,kanchana_calculated_2006,koide_effects_2017}.

In order to study the oxygen anions, we use microcrystals. Various techniques have been employed to grow cobalt- (CFO) and nickel-ferrite (NFO) films, with pulsed laser deposition (PLD) being a popular method for epitaxial layers \cite{opel_spintronic_2012}. But most of these films exhibit structural domains typically smaller than 100 nm due to defects that pin down the magnetic domains, resulting in unexpected properties such as high coercivity or new easy-axis behavior. In contrast, growth on Ru(0001) substrates using high-temperature oxygen-assisted molecular beam epitaxy (HOMBE)\cite{MontiPRB2012, LauraAdvMat2015, AnnaSciRep2018,SandraJCP2020} yields large spinel-oxide islands, typically originating from a single nucleation center, on top of a divalent oxide wetting layer. The islands can reach several micrometers in width and exhibit a flat top, occasionally down to the atomic level. Their thickness ranges from several nanometers to several tens of nanometers. In our case, due to the wide range of crystal sizes, both laterally and in terms of thickness, averaging techniques are inadequate for studying their magnetic properties. Instead, we employ x-ray photoemission electron microscopy (XPEEM)\cite{schneider_investigating_2002} combined with dichroic techniques to investigate the magnetic properties on individual islands.

We perform experimental measurements of oxygen x-ray magnetic circular dichroism in soft x-ray absorption spectromicroscopy, aimed to obtain magnetic domains maps in nickel and cobalt ferrite and XMCD spectra in order to estimate the dichroic component and the oxygen orbital magnetic moment through the sum rules applied to the oxygen K-edge. We combine the experimental results with electronic structure calculations based on the density functional theory (DFT) that have already demonstrated accuracy in the description of cubic spinel ferrites\cite{Tejera2021, Bernal2015}, addressing the specific conditions present in these particular samples. This way, we determine the influence of such conditions in the spin and orbital moments of all atoms, together with the electronic features contributing to the sum rules.

\section{Methods}

The growth of cobalt- and nickel ferrite crystals, as well as the subsequent x-ray absorption experiments, were conducted at the CIRCE end-station of the Alba Synchrotron Facility\cite{CIRCE}. It is equipped with
an instrument for performing low-energy electron microscopy and energy-filtered photoemission electron microscopy, providing a multitechnique approach to investigate the structural and chemical properties of surfaces.
This setup enables the acquisition of images with a spatial resolution of approximately 20 nm and an energy resolution down to 0.2 eV. The kinetic energy of the photoelectrons used to form the images can be selectively adjusted within this energy resolution.

After the growth of the samples, photoemitted electrons at low kinetic energies (typically 2 eV) were used for XAS and XMCD-PEEM. XAS spectra were extracted from selected areas on the surface using stacks of PEEM images collected at different photon energies. The x-ray beam was directed at the sample surface with a grazing angle of 16$^\circ$. Dichroic maps were obtained by measuring the asymmetry between two images acquired with circular left and right x-ray polarizations on a pixel-by-pixel basis. The XMCD spectra were extracted by integrating the intensity from a single domain region.

The synthesis of mixed nickel- and cobalt-iron oxides was performed on a ruthenium single crystal with (0001) orientation. The substrate was cleaned through multiple cycles of annealing in oxygen at 1200 K
under a molecular oxygen pressure of 10$^{-6}$ mbar, followed by flash heating to 1800 K under vacuum conditions. The growth of the oxides was carried out by co-depositing nickel or cobalt together with iron on the heated substrate (typically at 1100 K) in a molecular oxygen background pressure of 10$^{-6}$ mbar to obtain respectively iron-rich cobalt ferrite or iron-rich nickel ferrite. The deposition of nickel,
cobalt and iron was achieved using custom-made electron bombardment-heated evaporators.

The DFT calculations have been performed with the Vienna ab-initio simulation package (VASP)\cite{vasp} using the PAW method\cite{paw} and the PBE revised for solids (PBE$_{sol}$) parametrization of the GGA
exchange correlation functional. We considered both stoichiometric and Fe-rich Co and Ni cubic spinels, the latter simulated with a 28 atoms unit cell replacing two Co or Ni atoms by Fe in order to reproduce the measured experimental compositional ratio. Following previous calculations, a local Hubbard U term of 4 eV was added to the cation d bands based on the Dudarev approach, except for the direct Co spinel where a value of 6 eV is needed\cite{tesisCesar}. We also explored the effect of the Co or Ni distribution under Fe excess conditions, in terms of the relative interatomic distance between Co or Ni atoms. Our results evidenced both that the most stable configurations are those where Co-Co and Ni-Ni pairs are nearest neighbors, and that increasing their distance has only a limited influence on the values of the local magnetic moments under focus in this study. Thus we restrict the results presented here to the most stable configurations. All structures were fully relaxed until the forces on the atoms were below 0.01 eV/\AA$^2$.
Convergence in the total energies below 0.01 meV was guaranteed using an energy cutoff of 500 eV and sampling the Brillouin zone with a (10 $\times$ 10 $\times$ 7) k-mesh. The atomically resolved charges and spin moments were obtained after Bader analysis.

\section{Results and discussion}

\begin{figure}[htb]
	\centerline{\includegraphics[width=0.5\textwidth]{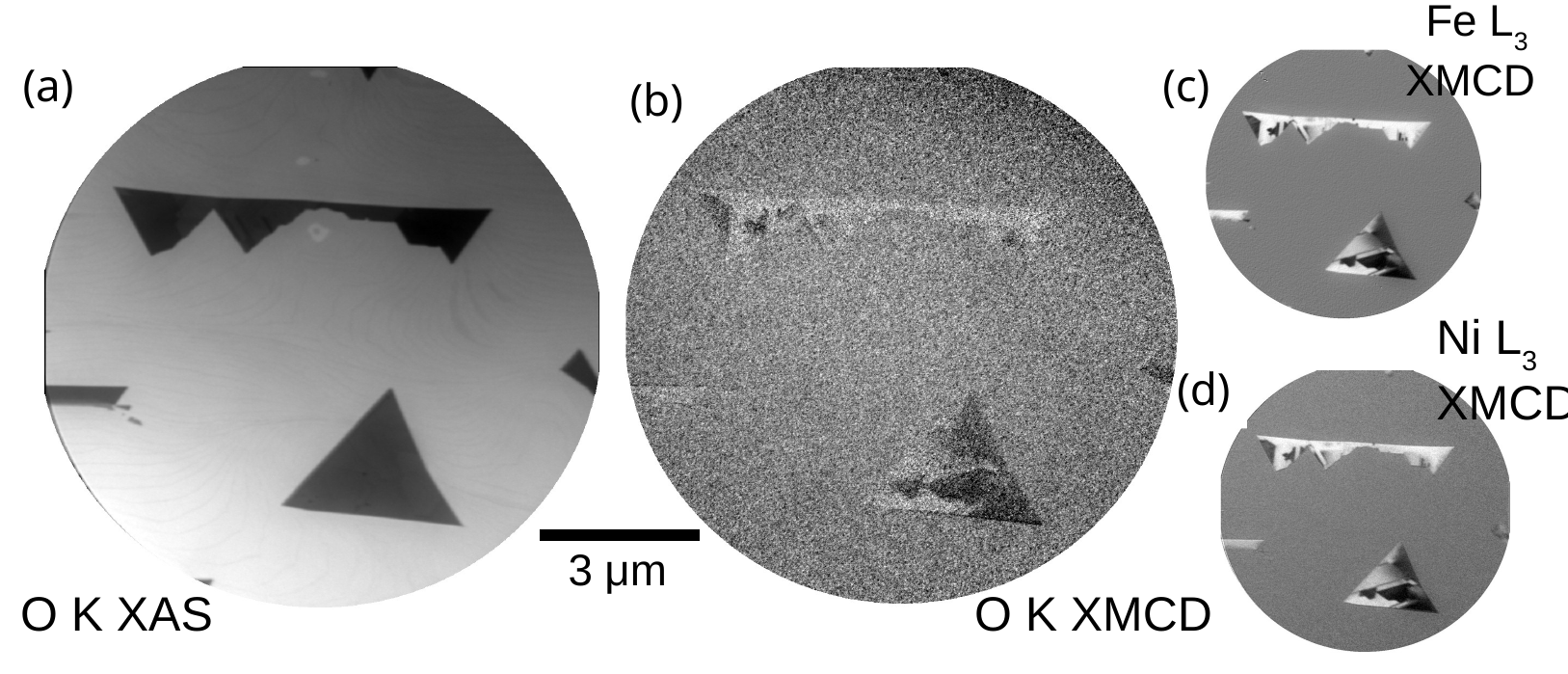}}
	\centerline{\includegraphics[width=0.5\textwidth]{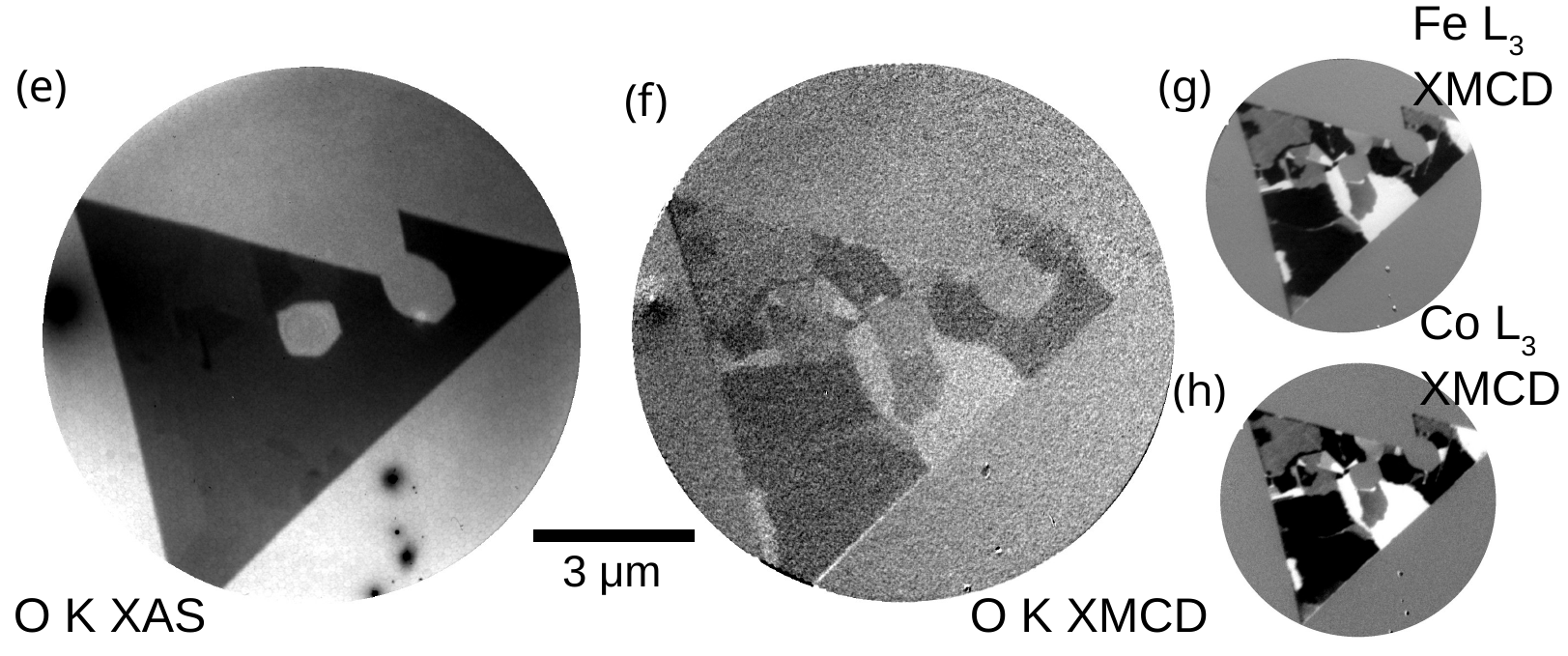}}
	\caption{PEEM images of nickel ferrite (a-d) and cobalt ferrite (e-h). (a) XAS PEEM image at the energy of the maximum intensity of the oxygen K-edge spectra for several nickel ferrite islands, (b) XMCD of oxygen at the same energy, (c) XMCD at the peak of the L$_3$ iron edge corresponding to the octahedral Fe$^{3+}$ cations, and (d) XMCD at the peak of the L$_3$ nickel edge corresponding to the Ni$^{2+}$ cations. (e) XAS PEEM image at oxygen for a cobalt ferrite islands, (b) XMCD of oxygen at the same energy, (c) XMCD at the peak of the L$_3$ iron edge corresponding to the octahedral Fe$^{3+}$ cations, and (d) XMCD at the peak of the L$_3$ cobalt edge corresponding to the Co$^{2+}$ cations. The latter two are adapted from Ref.~\citenum{LauraAdvMat2015}.}
	\label{fg:XASimages}
\end{figure}

We have previously reported the growth by high-temperature oxygen-assisted MBE on Ru(0001) of magnetite, cobalt ferrite and nickel ferrite\cite{SantosJPC2009, MatteoPRB2012, LauraPP2016, LauraAdvMat2015, SandraJCP2020, AnnaSciRep2018,SandraASS2022}. The behaviour is similar in all cases.

In this work, we focus on the most relevant aspects required for the oxygen characterization. We only summarize that the growth of these oxides begins
with the nucleation of islands composed of a divalent mixed oxide with a rock-salt structure, such as Fe$x$Co${1-x}$O\cite{SandraJCP2020} or Fe$x$Ni${1-x}$O\cite{AnnaSciRep2018}. These 2D islands continue to grow
until they form a complete layer, typically one or two atomic layers thick (the particular thickness depends on the oxygen background pressure\cite{IreneJPC2013,Lewandowski2021}), wetting the substrate.
Subsequently, three-dimensional spinel islands nucleate and grow on the surface. At high temperature, the islands remain well separated from each other, so they are presumably antiphase-boundary free, which gives rise to magnetic domains orders of magnitude larger than for growth by other methods.

The islands exhibit\cite{LauraAdvMat2015,AnnaSciRep2018} the spinel structure, as detected by low-energy electron diffraction as well as by x-ray absorption spectroscopy, with Ni$^{2+}$/Co$^{2+}$ and Fe$^{2+}$
ions predominantly occupying the octahedral (B) sites, while Fe$^{3+}$ ions are present in both octahedral (B) and tetrahedral (A) positions. The composition ratio of Ni/Co to Fe in the ternary oxide islands is
estimated to be close to 1:5 by measuring the edge-jump in area-selected X-ray absorption spectroscopy (XAS) spectra, so they are iron-rich ferrites.

Typical islands are shown in Figure~\ref{fg:XASimages}(a,e) for nickel ferrite and cobalt ferrite, respectively. Images were acquired at the maximum of the oxygen K-edge and are the result of averaging images obtained by illuminating with opposite circular polarization. The spinel islands correspond to the dark shapes in those images, arising from the work function differences between the spinel islands and the wetting layer with the rock-salt structure around the islands\cite{LauraAdvMat2015}. In addition, within the islands, contrast differences can be detected which are actually related to differences in thickness, as we confirmed by comparison with correlative atomic force microscopy\cite{LauraAdvMat2015,SandraASS2022}.

We have already studied the magnetic domains in such islands by measuring XMCD-PEEM at the L$_{3}$ absorption edges of the cations\cite{LauraAdvMat2015,AnnaSciRep2018}. We have observed large magnetic domains in remanence, determined their magnetic orientation and the origin of the observed domain wall pinning suggested\cite{SandraASS2022}. In Figure~\ref{fg:XASimages}(c,d,g,h) XMCD images show white regions corresponding to areas with local magnetization along the x-ray beam, black regions corresponding to areas with magnetization in the opposite direction, and gray regions corresponding to areas with either no local magnetization or magnetization orthogonal to the light direction. The domains detected at the edges of the two cations for each spinel (Ni/Fe and Co/Fe) have exactly the same orientation down to the smallest detectable details (compare images in Figure~\ref{fg:XASimages}(c,d) and (g,h)). This is expected since the signal arises from the cations in octahedral positions which are all coupled ferromagnetically.

\begin{figure}[htb]
	\centerline{\includegraphics[width=0.5\textwidth]{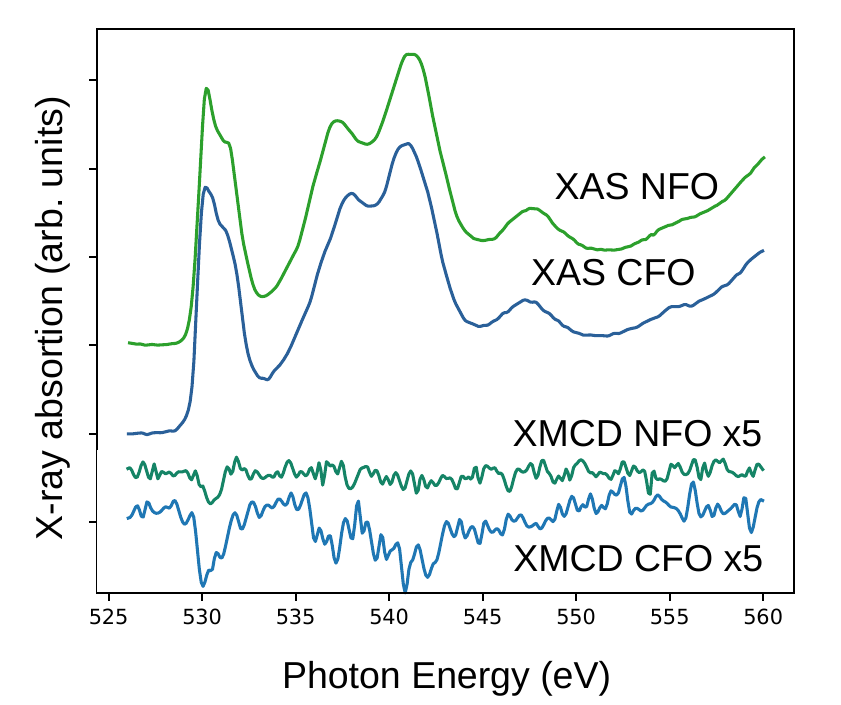}}
	\caption{Sum of the X-ray absorption spectra for opposite circular polarizations at the O K edge for 
the iron rich nickel ferrite islands (labelled NFO, green) and for the iron rich cobalt ferrite ones 
(labelled CFO, blue). The dichroic spectra (difference between the XAS spectra for two polarizations) are 
shown below (note the scale is expanded by a factor of $\times5$ in the latter).}
	\label{fg:XASspectra}
\end{figure}

In this work we focus on the contrast observed in the magnetic circular dichroism images acquired {\em at the oxygen K-edge}. These images show also black, white and gray domains, as displayed in Figure~\ref{fg:XASimages}(b,f). Comparing with the images obtained by conventional L$_{3}$ dichroism, these XMCD images show exactly the same features in the three elements: O, Fe, and Ni/Co. As in the Fe and Ni/Co edges the signal is magnetic, this implies that the dichroism observed in oxygen, acquired at the XAS edge from the 1s core level, is related to a magnetic origin. The contrast of the XMCD images is much smaller than for the cations. In our microcrystals, iron and cobalt have a typical XMCD contrast (difference divided by the sum) of about 24\% and nickel gives a smaller contrast (3\%). The oxygen signal is much smaller, 0.9\% for the cobalt ferrite and 0.5\% for the nickel ferrite.

We present full XAS and XMCD spectra around the oxygen K-edges for the nickel and cobalt ferrites in Figure~\ref{fg:XASspectra}. We first discuss the XAS spectra. The oxygen K-edge spectra of transition metal oxides\cite{de_groot_oxygen_1989,frati_oxygen_2020} shows a prominent double peak at 530~eV, a wider structure composed again of two peaks in the range of 535-545~eV, and another peak at higher energies, around 547~eV. The oxygen K-edge corresponds to the excitation from the oxygen 1s orbital to the unoccupied p-orbitals and the absorption peak structure closely reflects the unoccupied density of states of the oxygen anion\cite{de_groot_oxygen_1989}. This implies disregarding multiplet effects as well as the interaction of the core-level hole with the unoccupied density of states. In such terms, the first double peak observed (near 530~eV) reflects the hybridization of the oxygen 2p states with the d-bands of the cations. 
The O K‐edge doublet at 529.8 and 531.2 eV arises from 2p--3d hybridization, namely the $t_{2g}$-$e_g$ crystal‐field split in the spinel lattice. Because spinels host both octahedral  and tetrahedral  cations, the spectrum represents a convolution of two ligand‐field manifolds of opposite sign.  The other peaks are related to hybridization with s and p-derived bands of the cations\cite{frati_oxygen_2020}. In any case, the spectra of both spinel microcrystals are remarkably similar, containing the same features. We note that the expected main difference between the two materials is that nickel ferrite is expected to be more close to purely inverse in cation distribution, while the cobalt might be more mixed.

\begin{figure}[htb]
	\centerline{\includegraphics[width=0.5\textwidth]{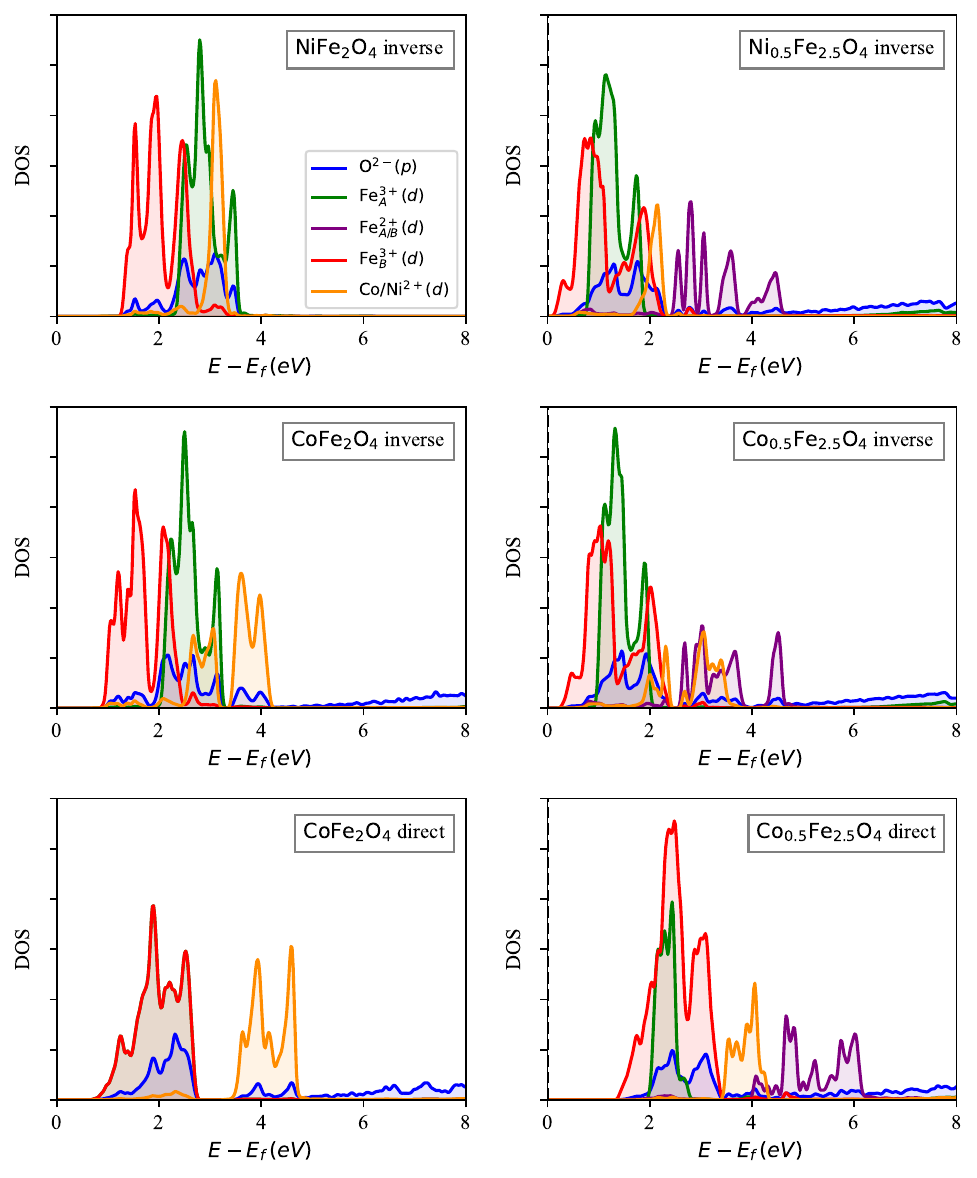}}
	\caption{DOS at the conduction band of stoichiometric (left) and Fe-rich (right) Ni and Co ferrites. 
The different colors correspond to contributions from different elements, separating A and B sites for Fe. 
	The energy zero is at the valence band edge (E$_f$).}
	\label{fg:DOS}
\end{figure}

Figure~\ref{fg:DOS} shows the theoretical unoccupied density of states (DOS), focusing on the energy range of the main peaks of the XAS spectra and resolving each atomic contribution. First we look at the left panels of the two upper rows, that refer to the inverse Co and Ni ferrites. Supporting the previous interpretation, the oxygen p band provides a narrow (width around 2 eV) double peak close to the valence band edge that overlaps mainly with the Fe$^{3+}_A$ and Co$^{2+}_B$ (or Ni$^{2+}_B$) d orbitals. Interestingly, as shown in the corresponding upper right panels of the figure, the existence of Fe excess shifts the bands closer to the valence band edge, narrowing and defining more clearly the double peak shape for O. The increase in the number of Fe$_B$ sites triggers the well-known charge disproportionation observed in magnetite at low temperatures, leading to two charge states (Fe$^{2+}_B$ and Fe$^{3+}_B$). As a consequence, there is a slight redistribution of the O band, resolving a first peak mainly overlapping with Fe$^{3+}_A$, and a second one overlapping with the cations at B sites.

We now turn to the XMCD spectra. We only detect significant dichroism around the position of the large double peak of the oxygen K-edge spectrum near 530~eV, corresponding to hybridization with the cation d-bands. This dichroism, which indicates the presence of an orbital moment, has also been observed in other oxides such as CrO$_{2}$ \cite{goering_direct_2002,huang_orbital_2002,kanchana_calculated_2006,koide_effects_2017}. The signal is quite noisy, as the data were acquired from a sub-micrometer sized area.

\begin{table}
	\begin{center}
		\begin{tabular}{ |c|c|c| }
 		\hline
 		Ni$_{0.5}$Fe$_{2.5}$O$_4$ & Co$_{0.5}$Fe$_{2.5}$O$_4$ \\
 		\hline
 		$1.6\times10^{-3} \mu_B$ & $4.4\times10^{-3} \mu_B$ \\
  		\hline
		\end{tabular}
	\end{center}
	\caption{Experimental orbital magnetic moments}
	\label{tb:exp_moments}
\end{table}

The dichroism in the L$_{3,2}$ cations has long been used to estimate their spin and magnetic moments. In the cations, the final-state orbital moment is considered to arise from spin-orbit interaction in the 3d band. This interaction is quite weak for 3d elements. Thus in pure 3d transition metals, the orbital moment is largely quenched, with values of the order of 0.10 $\mu_B$ for Fe, Co or Ni. For oxides in general and spinels in particular, larger orbital moments are observed specially in mixed spinels like cobalt ferrite and nickel ferrite, whose Co/Ni cations reside only (Ni) or mostly (Co) in octahedral sites. In such case, orbital magnetic moments of the order of up to 1~$\mu_B$ have been reported. Our own measurements for the microcrystals of spinels grown on Ru(0001) revealed in all cases a small orbital moment for Fe (typically of the order of 0.10$\mu_B$)\cite{SandraNano2018,LauraPhD2017}, while Ni in nickel ferrite has a slightly larger value of 0.2 $\mu_B$\cite{AnnaSciRep2018}, and the Co cations have a much larger orbital moment of 0.6 $\mu_B$\cite{LauraPhD2017}. This is in reasonable agreement with the expected larger orbital moment of Co cations in octahedral positions.

On the other hand, the interpretation of the dichroism at the K edge\cite{guo_interpretation_1998} relates the magnitude of the effect to the orbital magnetic moment of the element from which the 1s core level electrons are excited onto the p-projected states. The initial state, 1s, has spherical symmetric. In oxygen, the process does not involve directly the cations 3d bands as the final state is a p-band.

\begin{table*}
        \begin{center}
                \begin{tabular}{ |c|c|c|c|c|c| }
                \hline
                M$_s$/M$_l$ ($\mu_B$) & M /f.u. & Fe(B) & Fe(A) & Co or Ni & O\\
                CoFe$_{2}$O$_4$ & 3.00/0.29 & 4.11/0.02 & -3.96/-0.02 & 2.63/0.24 & 0.06/0.00 \\
                NiFe$_{2}$O$_4$ & 2.00/0.14 & 4.12/0.02 & -4.01/-0.02 & 1.58/0.16 & 0.06/0.00 \\
                Fe$_3$O$_4$ & 4.00/0.43 & 3.92/0.22 & -4.00/-0.02 & -- & 0.04/0.00 \\
                Co$_{0.5}$Fe$_{2.5}$O$_4$ & 3.50/0.16 & 3.96/0.07 & -4.02/-0.02 & 2.63/0.22 & 0.07/0.00 \\
                Ni$_{0.5}$Fe$_{2.5}$O$_4$ & 3.00/0.13 & 3.96/0.07 & -4.02/-0.02 & 1.59/0.15 & 0.07/0.00 \\
                \hline
                \end{tabular}
        \end{center}
        \caption{Calculated spin/orbital total magnetization per formula unit (M) and element specific average
magnetic moments for each magnetic sublattice at the inverse spinel structures, all in $\mu_B$.}
        \label{tb:dft_moments}
\end{table*}

The oxygen XMCD spectra for nickel and cobalt ferrite microcrystals are shown in Figure~\ref{fg:XASspectra}. The spectra are quite noisy, but the presence of dichroism is unambiguous as we have also confirmed it by comparing the XCMD between two domains with opposite magnetizations. The dichroic signal is clearly close to the first peak of the XAS spectra, and is about twice as large for cobalt ferrite compared to nickel ferrite. The XAS spectra on both spinel microcrystals shows an increasing background and does not reach a constant value beyond the edge in the range explored (up to 560 eV). Thus, the estimated orbital magnetic moments, presented in table~\ref{tb:exp_moments}, should be considered as order-of-magnitude values. The integral of the circular left and right polarized XAS spectra were obtained after subtracting an arctan-shaped background centered at 531~eV, and the number of holes was set to 0.5 following Ref.~\citenum{huang_orbital_2002}. The direction of the magnetic domains has been taken into account by the reconstructed surface vector magnetization obtained from several XMCD-PEEM images of the cations acquired at different azimuthal angles\cite{SandraNano2018,AnnaSciRep2018,SandraASS2022}. While the moments are small (and in the same range as those measured in CrO$_2$, for which $3\times10^{-3}$ was reported\cite{huang_orbital_2002}), they are consistent with the XMCD images.

The DFT calculations also enable to determine element and site specific values for the spin and orbital moments. They are shown in table~\ref{tb:dft_moments}, both for the stoichiometric and Fe-rich forms. As a reference, values for magnetite are provided too. The atomic moments in the table correspond to the average contribution at the unit cell for each chemical species. We should remark that, while the cation spin moments have low dispersion (except that linked to two Fe$_B$ valence states when charge disproportionation takes place under Fe excess conditions), the opposite occurs for O. This can be understood as the O magnetization being induced by hybridization to the cations, and different O atoms are coordinated to different cationic species. Importantly, in spite of the dispersion, the O spin moments are always parallel to the magnetization of the B sublattice, explaining the observation of uniform magnetic domain contrast at the XMCD images in Figure~\ref{fg:XASimages}. Comparing now the spin magnetizations of the stoichiometric ferrites, the larger magnetic moment of Co as compared to Ni causes a larger net magnetization per formula unit, but there are few variations among both materials regarding the Fe and O contributions. The same holds true for the orbital moments, where in addition the largest component at each system comes respectively from Co or Ni, in good agreement with the experimental observations. However, while the Co orbital moment is much larger than that of Ni, and both are over the Fe contribution at Co and Ni ferrites, in magnetite the theoretical orbital Fe$_B$ moment is similar to that of Co. This discrepancy with the experimental observations needs further study, but probably is related to the existence of several Fe charge states in magnetite.

Finally, we remark that in all cases the orbital moment of O is null, reinforcing the conclusion that the XMCD signal should be interpreted as an indirect measurement of the cation contributions. Consideration of the actual Fe excess in the experimental samples does not alter this scenario, as demonstrated by the values at the last two rows of table~\ref{tb:dft_moments}. The slight reduction in the cation orbital contribution coming from the supression of some Co or Ni atoms is somehow compensated by the increase of the orbital moment of Fe$_B$ cations in such cases, that reaches values of 0.12 $\mu_B$ at some Fe$_B$ sites. The overall effect of Fe excess is an increase of the net spin magnetization (more noticeable at Ni ferrite) and a decrease of the orbital component (more pronounced at Co ferrite), but with negligible variations in the element specific contributions from Co, Ni and O.

As mentioned in the introduction, there is some probability for Co cations to occupy tetrahedral coordination sites. We have explored the impact of the Fe excess conditions in the tendency of Co cations to occupy A or B sites, 
the results are in table~\ref{tb:directE} also considering Ni ferrite. The table provides the energy difference between the inverse and direct spinel forms at each stoichiometry, placing the energy zero at the most stable structure. Not surprinsingly, at bulk stoichiometries the inverse form is preferred for both oxides, with a larger
energy barrier for Ni than for Co. At Fe-rich conditions, while Ni continues following this trend, for Co the direct spinel becomes more stable than the inverse structure. A mixed form with Co at both A and B sites seems less stable, but the energy barriers between the different configurations are low, admitting their coexistence. Though not shown, our calculations indicate that the stability of the direct structure over the inverse one at Co$_{0.5}$Fe$_{2.5}$O$_4$ is independent of the interatomic distance between Co cations.
These results indicate the convenience to evaluate the orbital moments at the direct Co ferrite, both under stoichiometric and Fe excess conditions.

Before addressing this, we remind that the stabilization of the direct form has implications for the magnetic anisotropy. The inverse Co ferrite is a hard magnet where the easy axis depends on the specific Co distribution\cite{fritsch_epitaxial_2010}, while magnetite, Ni ferrite and the direct Co spinel are soft magnets. The presence of direct CoFe$_2$O$_4$ thus reduces the magnetic hardness, supporting the adequacy of the XMCD experimental setup to capture the largest contribution to the magnetization. The XMCD measurements are performed with the X-ray beam at
grazing incidence, thus they are mainly sensitive to the in-plane component of the magnetization. In soft magnets it is expected that the shape anisotropy contribution dominates to orient the magnetization in-plane in the near surface region, reducing in this way the stray field. On the contrary, for the hard inverse Co ferrite the easy axis depends on the specific Co
distribution\cite{fritsch_epitaxial_2010} and can add an out-of-plane magnetization component, as been previously reported in the cobalt ferrite microcrystals\cite{SandraASS2022}.

\begin{table}
        \begin{center}
                \begin{tabular}{ |c|c|c|c| }
                \hline
                X &  Struc & XFe$_2$O$_4$ & X$_{0.5}$Fe$_{2.5}$O$_4$  \\
                \hline
                Co & {\it I} &  0.00      &   0.04   \\
                Co & {\it D} &  2.13      &   0.00   \\
                Co & {\it M} &   --       &   0.37  \\
                \hline
                Ni & {\it I} &  0.00      &   0.00  \\
                Ni & {\it D} &  5.18      &   8.24  \\
                \hline
                \end{tabular}
        \end{center}
        \caption{Total energy differences in eV/unit cell, referred to the most stable configuration, between 
the direct and inverse forms of both stoichiometric and Fe-rich Co and Ni ferrites. {\it I} accounts for 
inverse, {\it D} for direct and {\it M} for the mixed structure with Co$_A$ and Co$_B$.}
        \label{tb:directE}
\end{table}

Coming now to the magnetic moments, table~\ref{tb:directM} summarizes their values at  CoFe$_2$O$_4$ and Co$_{0.5}$Fe$_{2.5}$O$_4$ when all Co atoms at the unit cell (cases termed Direct in the table) or only half of them (Mix) occupy A positions. In general, the presence of Co$_A$ atoms instead of Fe$_A$ serves to enhance the net magnetization, as it arises from the difference between the opposite contributions of the B and A sublattices, and Co has a lower spin moment than Fe.
Co$_A$ has a slightly lower spin contribution than Co$_B$, and in general there is a moderate increase of the O and Fe spin contributions as compared to the inverse spinel. But the most noticeable changes occur at the orbital components. At CoFe$_2$O$_4$, Co continues to provide the largest orbital contribution among all cations, and this component aligns parallel to the spin one, resulting in a net negative orbital magnetization, opposite to the experimental observation. The Fe orbital moment is positive but low, and the larger number of Fe cations over Co serves to reduce the absolute value of the negative orbital magnetization. Anyhow, the O orbital moment continues to be zero. This discrepancy in the sign of the orbital contribution is somehow solved at Co$_{0.5}$Fe$_{2.5}$O$_4$, where the situation is more complex: the combination of Co$_A$ and Fe$_A$ promotes the charge disproportionation at the Fe$_B$ sublattice similarly to magnetite, leading to high positive values of the orbital Fe moment at Bi sites. The O orbital moment continues to be zero, but if we have a look at the unoccupied DOS in these direct spinels (see bottom panel of Figure \ref{fg:DOS}), there is a significant overlap of the O p band with the Fe$_B$ states. In the case of the mixed spinel with both Co$_A$ and Co$_B$, as seen in Table \ref{tb:directM} the Co sublattices provide opposite contributions, resulting in a reduction of the net orbital magnetization. In any case, the local Co$_B$ and O orbital moments are similar to the inverse ferrite. In summary, our main conclusion is that, even though the presence of Co cations at A sites cannot be discarded
and they will reduce the overall orbital magnetization, this will not add any contribution to the O orbital moment, while keeping unquenched values at the cation sublattices.

\begin{table*}
	\begin{center}
		\begin{tabular}{ |c|c|c|c|c|c| } 
 		\hline
                M$_s$/M$_l$ ($\mu_B$) & M /f.u. & Fe(B) & Fe(A) & Co & O\\
		D CoFe$_{2}$O$_4$ & 7.00/-0.15 & 4.24/0.02 & --         & -2.53/-0.20 & 0.26/0.00 \\
		D Co$_{0.5}$Fe$_{2.5}$O$_4$ & 5.50/0.02 & 4.13/0.12 & -4.17/-0.02 & -2.69/-0.21 & 0.17/0.00 \\
		M Co$_{0.5}$Fe$_{2.5}$O$_4$ & 3.00/0.02 & 3.58/0.05 & -4.19/-0.02 & $\pm 2.71$/$\pm 0.22$ & -0.07/0.00 \\
  		\hline
		\end{tabular}
	\end{center}
        \caption{Same as table\protect~\ref{tb:dft_moments} for the direct (D) and mixed (M) forms of 
stoichiometric and Fe-rich Co spinel ferrites. The double sign of the Co contribution at M 
Co$_{0.5}$Fe$_{2.5}$O$_4$ corresponds to Co$_B$ (positive) and Co$_A$ (negative) atoms.}
	\label{tb:directM}
\end{table*}

Thus in all cases, the DFT calculations indicate that an average spin moment of the order of $10^{-2}$ $\mu_B$ should be induced in the oxygen anions by the hybridization with the cations, and a slightly higher one for iron-rich spinels when compared with stoichiometric ones. While no orbital moment is detected in the calculations, the error limit of the DFT calculations would be in the range of the experimental measurements. Even then, for oxygen we do not expect that the observed experimental orbital moment of 10\% of the predicted spin moment can arise from spin-orbit coupling in an element as light as oxygen.

However, other oxygen K-edge XMCD experiments have obtained values on the order of $10^{-3}$ $\mu_B$ per O. A similar discrepancy in CrO$_2$ has been traced to the fact that the 1s$\rightarrow$2p transition probes unoccupied O p states that are strongly hybridized with cation 3d orbitals, in agreement with our calculations for the spinels. Because those cation d-states carry significant spin–orbit coupling, they impart orbital polarization onto the mixed O p–M d bands. Consequently, the O K-edge XMCD signal reflects the cation-driven orbital moment in the final state, not an intrinsic orbital moment on oxygen itself.

\section{Conclusions}

We have observed the x-ray magnetic circular dichroism at the oxygen K-edge of spinel microcrystals grown by high-temperature oxygen assisted molecular beam epitaxy. The dichroism allows to observe the magnetic domains of the spinel islands, which reproduce the domains observed using the L$_3$-edge dichroism from the cations. The XMCD spectra show significant dichroism at the energy ranges where there is significant hybridization with both the d-derived and sp-derived bands of the cations. Using the sum rules, we obtain values of the "nominal" orbital magnetic moment of the oxygen of  $1.6\times10^{-3}\mu_B$ for the iron-rich nickel ferrite and $4.4\times10^{-3}\mu_B$ for the iron-rich cobalt ferrite islands. First-principles calculations provide significant values of the orbital moments only at the cations, mainly Co and Ni, but not for the O atoms. Fe excess conditions modify only slightly the local moments
under inverse geometries, but favor the presence of direct Co ferrite forms. This has a significant influence on the Co and Fe orbital moment values but does not add any contribution to the O orbital moment. DFT oxygen spin moments are obtained in the $10^{-2}$ $\mu_B$ range, suggesting that orbital moments should be negligible. However, in all cases the calculations indicate the existence of hybridization between the O p-band and the cation d-bands in the unoccupied energy region contributing to the spectra, particularly relevant for those cations providing the largest orbital components. We thus suggest that the oxygen K-edge dichroism is not actually sampling the atomic localized orbital moment of oxygen but rather a combined effect of hybridized bands of the oxygen and the cations.

\section*{Acknowledgements}

This work is supported by the Grants PID2021-124585NB-C31/-C33, PID2021-122980OB-C55, PID2020-117024GB-C43 and TED2021-130957B-C53/-C54 funded by MCIN/AEI/10.13039/501100011033 and by “ERDF A way of making Europe” and by the "European Union NextGenerationEU/PRTR". P. N. acknowledges the SciMat Priority Research Area budget under the program Excellence Initiative - Research University at the Jagiellonian University in Kraków - Grant No. 75.1920.2021These experiments were performed at the CIRCE beamline of the ALBA Synchrotron Light Facility. V.S. acknowledges financial support from PIPF-2023/ECO-30425.

\bibliography{OM}

\end{document}